\documentclass[12pt]{article}
\usepackage{amsmath}
\usepackage{multirow}
\usepackage{amsfonts}
\usepackage{amssymb}
\def\hybrid{\topmargin -20pt    \oddsidemargin 0pt
        \headheight 0pt \headsep 0pt
        \textwidth 6.25in       
        \textheight 9.5in       
        \marginparwidth .875in
        \parskip 5pt plus 1pt   \jot = 1.5ex}
\hybrid
\numberwithin{equation}{section}
\numberwithin{table}{section}
\newcommand{\beq}{\begin{equation}}
\newcommand{\eeq}{\end{equation}}
\newcommand{\bi}{\begin{itemize}}
\newcommand{\ei}{\end{itemize}}
\newcommand{\bea}{\begin{eqnarray}}
\newcommand{\eea}{\end{eqnarray}}
\newcommand{\ba}{\begin{array}}
\newcommand{\ea}{\end{array}}
\newcommand{\bt}{\begin{tabular}}
\newcommand{\et}{\end{tabular}}
\newcommand{\bc}{\begin{center}}
\newcommand{\ec}{\end{center}}

\newcommand{\Kh}{{\hat{K}}}

\newcommand{\Ah}{{\hat{A}}}
\newcommand{\Bh}{{\hat{B}}}

\newcommand{\nn}{\nonumber}

\newcommand{\IM}{\textrm{Im} \,}
\newcommand{\RE}{\textrm{Re} \,}

\newcommand{\cref}{{\bf [check ref]}}

\begin{document}

\begin{titlepage}
\begin{center}

\hfill hep-th/yymmddd\\
\vskip 0.3cm

\vskip 2cm

{\Large \bf Massive Tensor Multiplets in $N=1$ Supersymmetry}\footnote{%
Work supported by: DFG -- The German Science Foundation,
European RTN Program HPRN-CT-2000-00148 and the
DAAD -- the German Academic Exchange Service.}\\

\vskip 2cm

{\bf Jan Louis and Waldemar Schulgin}  \\
\vskip 0.3cm

{\em II. Institut f{\"u}r Theoretische Physik\\
Universit{\"a}t Hamburg\\
Luruper Chaussee 149\\
 D-22761 Hamburg, Germany}\\[1ex]
 {\tt jan.louis@desy.de, waldemar.schulgin@desy.de} \\

\end{center}

\vskip 2cm

\begin{center} {\bf ABSTRACT } \end{center}

\noindent
We derive the action for a massive tensor multiplet coupled
to chiral and vector multiplets 
as it can appear in orientifold compactifications of type IIB string theory. 
We compute the potential of the theory and show its consistency with
the corresponding Kaluza-Klein reduction of 
$N=1$ orientifold compactifications. 
The potential contains an explicit mass term for the scalar 
in the tensor multiplet which does not arise from eliminating 
an auxiliary field.
A dual action with an additional massive vector multiplet
is derived at the level of superfields.
\vfill


October 2004\\

\end{titlepage}

\section{Introduction}
Consistent string theories as we know them today come equipped with some
amount of supersymmetry. As a consequence there is a continuous
vacuum degeneracy parameterized by the vacuum expectation values
of scalar fields called moduli. In order to make contact 
with particle physics it is necessary to break the supersymmetry 
and lift the vacuum degeneracy. It has been realized that
compactifications of string theory with non-trivial background fluxes 
generically generate a potential for the moduli scalars and spontaneously break
supersymmetry \cite{Bachas,PS,TV,Mayr,GKP,CKLT} possibly 
with a stable local de Sitter ground state \cite{KKLT,BKQ}.

In the low energy supergravity
the fluxes appear as gauge or mass parameters and turn an ordinary
supergravity into a gauged or massive supergravity.
In the latter case the fluxes induce mass terms for some of the 
$p$-form gauge potentials which are present in string theories
\cite{Romans,HLS,LM}. 
For example, in $N=2$ theories in four space-time dimensions 
massive two-forms appear in compactification of
type II string theory on Calabi-Yau threefolds when both 
electric and magnetic three-form fluxes are turned on \cite{LM}.
The corresponding $N=2$ supergravity theories have been constructed 
recently in \cite{vaula}. 
The same phenomenon can also be observed in $N=1$ 
compactifications of type IIB on Calabi-Yau orientifolds
with $O5$- or $O9$-planes \cite{TGL}. 
Surprisingly, in terms of the 
low energy $N=1$ effective supergravity  a massive two-form
has not been discussed in considerable detail in the literature
\cite{Siegel,GGRS,BK}.\footnote{%
While this manuscript was being
prepared we received a preprint \cite{DF} which discusses the same 
topic. Some phenomenological applications have recently been discussed
in \cite{Kors}.}
 The purpose of this paper
is to (partially) close this gap and study the superspace 
action proposed in \cite{TGL}.

In $N=1$ supersymmetry an antisymmetric tensor resides in a 
chiral spinor superfield $\Phi_\alpha$ while its three-form 
field strength is a member of a linear Multiplet $L$ \cite{Siegel}.
When the antisymmetric tensor is massless the action is expressed 
in terms of $L$ only and well known \cite{Siegel,GGRS,BK,Gates-pformen,LR,Quevedo,BGG}. 
However, when the 
antisymmetric tensor is massive the action explicitly features 
$\Phi_\alpha$ and gauge invariance requires a
St\"uckelberg-type coupling  to at least one vector multiplet $V$. 
In this paper we derive the supersymmetric action for such a 
massive antisymmetric tensor coupled to $n_V$
vector multiplets $V^A,\, A=1,\ldots,n_V$.
We additionally allow for the possibility
that the mass depends on an arbitrary number of chiral multiplets
$N^i,\, i=1,\ldots,n_c$. This situation is partly motivated by the results
of the orientifolds compactifications and
we explicitly rederive the supersymmetric effective action obtained  
in \cite{TGL} from a Kaluza-Klein reduction.
Curiously the potential is not determined solely by the auxiliary fields
but contains an explicit mass term for the scalar in the tensor multiplet.

This paper is organized as follows.
In section~2.1 we discuss the spinor superfield $\Phi_\alpha$, 
its field 
strength $L$ and their gauge transformation properties.
In section~2.2 we write down the most general gauge invariant 
superspace action for $\Phi_\alpha$
including a St\"uckelberg-type mass term and explicitly give
its corresponding
component form. We discuss in detail the resulting potential 
and show that it does not have a `standard' $N=1$ form.
In section~3 we perform the duality transformation in superfields
and rewrite the action in terms of 
$n_V-1$ massless and one massive vector multiplet. 
An expanded version of this work can be found in \cite{schulgin}.

\section{The action of a massive antisymmetric tensor}
\subsection{The supermultiplets}\label{superm}
Let us start by introducing the supermultiplets which contain an antisymmetric
tensor $B_{mn}$ and its field strength $H_{mnp}$. The latter resides in 
a linear superfield $L$ which is real and obeys the additional constraint
\beq\label{constraint}
{D}^2 L\ =\ \overline{D}^2 L\ =\ 0\ ,
\eeq
where $D_\alpha$ is the superspace covariant derivative.\footnote{We
are following the conventions of \cite{WB} and abbreviate  
${D}^2={D}^{\alpha}{D}_{\alpha}$ and 
${\overline{D}}^2={\overline{D}}_{\dot{\alpha}}{\overline{D}}^{\dot{\alpha}}$.}
In components it contains a real scalar $C$, the field strength
of an antisymmetric tensor $H_{mnp} = \partial_{[m}B_{np]}$
and a Weyl fermion $\eta$. Its $\theta$-expansion is obtained by solving
\eqref{constraint} and reads
\begin{equation}\label{linearsuperfield}
L=C+\theta\eta+\bar{\theta}\bar{\eta}
+\frac12 \theta\sigma^{m}\bar{\theta} \epsilon_{mnpq}H^{npq}
-\frac{i}{2}\theta\theta\bar{\theta}\bar{\sigma}^m\partial_m\eta
-\frac{i}{2}\bar{\theta}\bar{\theta}\theta\sigma^m\partial_m\bar{\eta}
-\frac{1}{4}\theta\theta\bar{\theta}\bar{\theta}\Box C \ .
\end{equation}
The antisymmetric tensor $B_{mn}$ itself is contained in a
chiral spinor multiplet $\Phi_\alpha$ defined via 
\begin{equation} \label{Beziehung.L.Phi}
L\ =\ \frac{1}{2}(D^{\alpha}\Phi_{\alpha}
+\overline{{D}}_{\dot{\alpha}}\overline{\Phi}^{\dot{\alpha}})\ ,
\qquad \overline D_{\dot\beta} \Phi_\alpha = 0\ .
\end{equation}
With this definition 
the constraints \eqref{constraint} are solved identically.
Furthermore, due to the identity 
$D^{\alpha}\bar{D}^2D_{\alpha}=\bar{D}_{\dot{\alpha}}D^2\bar{D}^{\dot{\alpha}}$,
$L$ is left invariant by the gauge transformation
\begin{equation} \label{Eichung}
\Phi_{\alpha} \rightarrow \Phi_{\alpha} 
+\frac{i}{8}\, \overline{{D}}^2{D}_{\alpha}\Lambda\ ,
\end{equation} 
where $\Lambda$ is a real superfield. 

$\Phi_{\alpha}$ contains off shell eight bosonic plus eight fermionic
degrees of freedom, four of which are rendered unphysical
by the gauge invariance \eqref{Eichung}. This can bee seen as follows.
A chiral spinor superfield $\Phi_\alpha$ has a generic
$\theta$-expansion 
\begin{equation}\label{Phiexp}
\begin{split}
\Phi_{\alpha}=&\chi_{\alpha}-\Big(\frac{{\delta_{\alpha}}^{\gamma}(C+iE)}{2}
+\frac{1}{4}{(\sigma^m\bar{\sigma}^n)_{\alpha}}^{\gamma}B_{mn}\Big)\theta_{\gamma}
+\theta\theta\left(\eta_{\alpha}+i{\sigma_{\alpha\dot{\alpha}}}^m\partial_m\bar{\chi}^{\dot{\alpha}}\right) \ .
\end{split}
\end{equation}
Since $\overline D ^2 D_ \alpha \Lambda$ is chiral it enjoys the expansion
\beq\label{Lambdaexp}
\frac{i}{8} \bar D^2 D_ \alpha \Lambda = -\frac12 \lambda_\alpha
-\Big({\delta_\alpha}^\beta \frac{iD}{2}
+\frac14 {(\sigma^m \bar\sigma^n)_\alpha}^\beta (\partial_m\Lambda_n-\partial_n\Lambda_m\Big)\theta_\beta -\frac{i}{2}\theta\theta \sigma_{\alpha\dot{\alpha}}^m\partial_m\bar \lambda ^{\dot {\alpha}} \ .
\eeq
Comparing \eqref{Phiexp} and \eqref{Lambdaexp} we see that 
the component fields $\chi_\alpha$ and $ E$ of $\Phi_\alpha$ can be gauged
away leaving only
the physical degrees of freedom which are $C, B_{mn}$ and $\eta$
in the component expansion 
of $\Phi_{\alpha}$ 
\begin{equation}\label{Phiexpansion}
\Phi_{\alpha}=
-\frac12\, \theta_{\gamma}\big({\delta_{\alpha}}^{\gamma}\, C
+\frac{1}{2}{(\sigma^m\bar{\sigma}^n)_{\alpha}}^{\gamma}\, B_{mn}\big)
+\theta\theta\, \eta_{\alpha}\ .
\end{equation}
In this gauge the left over gauge invariance is the 
standard two-form gauge invariance
\begin{equation}\label{gaugei}
B_{mn}\rightarrow 
B_{mn}+\partial_{m}\Lambda_{n}-\partial_{n}\Lambda_{m}\ ,\qquad
\eta\rightarrow \eta\ ,\qquad
C\rightarrow C\ .
\end{equation}

\subsection{Action of a massive tensor multiplet}
The next step is the construction of a gauge invariant Lagrangian 
for a massive antisymmetric tensor. Let us start with the kinetic term
which arises from \cite{BGG}
\beq\label{Lkin}
{\cal L}_{\rm kin} = -\int d^2\theta d^2\bar{\theta}\, K(L)\ ,
\eeq
where $K(L)$ is an arbitrary real function.
Using \eqref{linearsuperfield} one finds the component action
\bea\label{Lkincomponent}
{\cal L}_{\rm kin} &=&
-\frac{1}{4}K^{\prime\prime}\,\Big(\partial_m C\partial^m
C+i(\eta\sigma^m\partial_m\bar{\eta}
+\bar{\eta}\bar{\sigma}^m\partial_m\eta)
+\frac{3}{2}H_{mnp}H^{mnp}\Big)\nonumber\\
 &&
-\frac{1}{8}\,K^{\prime\prime\prime}\,\eta\sigma^m\bar{\eta}\epsilon_{mnpq}H^{npq}
-\frac{1}{48}\,K^{\prime\prime\prime\prime}\,\eta\eta\bar{\eta}\bar{\eta}\ ,
\eea
where $K^{\prime\prime}= \partial_C^2 K ,\ K^{\prime\prime\prime} = \partial_C^3 K $ etc. .

The mass term for $B_{mn}$ arises from a chiral superspace integral
which is quadratic in $\Phi_\alpha$ and reads
$\mathcal{L}_{\rm m} \sim \int d^2\theta\,
\Phi^{\alpha}\Phi_{\alpha} + {\rm h.c.\,}$. 
Taken at face value such a mass term is not gauge invariant
under the gauge transformation (\ref{Eichung}).
However, by appropriately coupling 
the spinor superfield $\Phi_\alpha$
to $n_V$ Abelian vector multiplets $V^A,\, A=1,\ldots,n_V$ 
gauge invariant mass terms can be constructed. The couplings have to be  such
that the vector multiplets
provide the necessary 
degrees of freedom for a massive $B_{mn}$ to exist. Or in other words we need
to employ a St\"uckelberg mechanism for $B_{mn}$ \cite{stuck}.

The chiral field strengths of the vector multiplets are denoted by
$W_\alpha^A = -\frac14 \overline D^2 D_\alpha V^A$ and 
have a component expansion \cite{WB}
\begin{equation}\label{Wexpansion}
W_{\alpha}^A=
-i\lambda_{\alpha}^A
+\big({\delta_{\alpha}}^{\beta}D^A
-\frac{i}{2}{(\sigma^m\bar{\sigma}^n)_{\alpha}}^{\beta}F_{mn}^A\big)\theta_{\beta}
+\theta\theta{\sigma_{\alpha\dot{\alpha}}}^m
\partial_m\bar{\lambda}^{\dot{\alpha}A}\ ,
\end{equation}
where $F_{mn}^A = \partial_mv_n^A-\partial_nv_m^A$ are the field
 strengths of $n_V$ 
$U(1)$ gauge boson $v_n^A$.
{}From the component form of the gauge transformation \eqref{gaugei}
we see that $\check F_{mn}^A := F_{mn}^A - m^A B_{mn}$ 
are gauge invariant combinations provided we assign 
to the gauge bosons $v^A_n$ the transformation laws
\beq\label{vtrans}
v^A_n \to v^A_n + m^A \Lambda_n\ .
\eeq
At this point the $m^A$ are $n_V$ parameters which, 
as we will see shortly, are related to the mass of $B_{mn}$.
In terms of superfields the transformation law \eqref{vtrans}
translates into\footnote{Of course one also has the standard $U(1)$
 gauge invariance 
$V^A \to V^A + \Sigma^A +\overline\Sigma^A,\, W_\alpha^A\to
 W_\alpha^A$
for chiral superfield parameters $\Sigma^A$.}
\beq\label{Vtrans}
V^A  \to V^A + m^A \Lambda \ , \qquad
W^A_\alpha  \to  W^A_\alpha 
-\frac{1}{4}m^A\overline{{D}}^2 {D}_\alpha\Lambda\ .
\eeq
Physically the transformation laws \eqref{Eichung},
\eqref{vtrans} and  \eqref{Vtrans} 
state that one vector superfield is `pure gauge' or in other
words can be `eaten' by the spinor superfield $\Phi_\alpha$.
As we will see shortly one can remove one of
the vector multiplets from the action
by going to the equivalent of the unitary gauge. However, it turns out to be
 more convenient to keep all vectors and leave the action
 manifestly gauge invariant under the combined transformation 
\eqref{Eichung} and \eqref{Vtrans}. 
Such an invariant action we are going to construct next.

We need to find all Lorentz-invariant expressions which are also 
singlets under the 
combined transformations given in \eqref{Vtrans} and \eqref{Eichung}.
One immediately
observes that $(W^A_\alpha-2im^A\Phi_\alpha)$ 
is a gauge invariant combination
of superfields. In order to find possible other combinations
we make the Lorentz-invariant Ansatz 
\begin{equation}
\mathcal{L}_{\rm m}
=\frac14 \int d^2\theta\left(f_{AB}W^AW^B + g_A\,\Phi W^A +
h\, \Phi\Phi \right)\ +\ {\rm h.c.} \ ,
\end{equation}
and determine the couplings $f_{AB},g_A$ and $h$ by 
demanding $\mathcal{L}_{\rm m}$ to be invariant under
\eqref{Vtrans} and \eqref{Eichung} up to a total derivative.
This determines 
\beq\label{lagrangian}
\mathcal{L}_{\rm m}=
\\ 
\frac{1}{4}\int\!\! d^2 \theta\Big(
f_{AB}(N)(W^A-2im^A\Phi)(W^B-2im^B\Phi)
+ 2 e_A\Phi(W^A-im^A\Phi)\Big)\ +\ {\rm h.c.}\ .
\eeq
The first term in the Lagrangian \eqref{lagrangian} contains the 
manifestly gauge invariant combination $(W^A_\alpha-2im^A\Phi_\alpha)$
and thus the  arbitrary coupling matrix 
$f_{AB}(N^i)$ can depend holomorphically on  
a set of $n_c$ chiral
superfields $N^i, i = 1, \ldots, n_c$.\footnote{Coupling the 
spinor superfield to chiral multiplets also requires a kinetic term for the 
$N^i$ but since this term plays no role for the purpose of this paper
we do not include it here.}
 The second term in \eqref{lagrangian} is only gauge invariant
up to a total derivative and thus its couplings $e_A$ have to be constants.
The Lagrangian \eqref{lagrangian} is the first result of this paper
in that, together with \eqref{Lkin},
it gives the most general Lagrangian of a massive spinor multiplet
coupled to $n_V$ vector and $n_c$ chiral multiplets. 
This form of the action has also been proposed in \cite{TGL}.

Before we expand the action in components 
let us briefly discuss the limit $m^A =0$ where 
\eqref{lagrangian}
reduces to $\mathcal{L}_{\rm m}=\tfrac14\int d^2\theta\,
(f_{AB}W^AW^B + 2e_A\Phi W^A) +{\rm h.c.}\,$. The second term 
is known as a four-dimensional Green-Schwarz term 
and it is 
conventionally written in terms of a full superspace 
integral including $L$ but not $\Phi_\alpha$ \cite{Quevedo,BGG} . 
Indeed, using \eqref{Beziehung.L.Phi}, the definition of $W_\alpha$
and partial integration one shows
\bea\label{GS}
e_A\int d^2\theta\, \Phi W^A + {\rm h.c.}
&=&-\frac{1}{4}\,e_A\int d^2\theta\,\Phi\overline{{D}}^2{D}V^A
-\frac{1}{4}\,e_A\int d^2\bar\theta\,
\bar{\Phi}{D}^2\overline{{D}}V^A\nonumber\\
 & =&e_A\int d^2\theta d^2\bar\theta\,
\left({D}\Phi+\overline{{D}}\overline{\Phi}\right)V^A
=2e_A\int d^2\theta d^2\bar\theta\, LV^A\ .\nonumber
\eea

Let us now expand the action \eqref{lagrangian} in components.
Apart from using \eqref{linearsuperfield}, \eqref{Phiexpansion},
\eqref{Wexpansion} we also need 
\bea \label{Nexpansion}
N^i&=&A^i+\sqrt{2}\theta\chi^i+\theta\theta F^i\ ,\\
f_{AB}(N)&=&f_{AB}(A)+\sqrt{2}\theta\chi^i\partial_if_{AB}
+\theta\theta\big(F^i\partial_if_{AB}
-\frac{1}{2}\chi^i\chi^j\partial_i\partial_jf_{AB}\big)\ . \nonumber
\eea
Inserting \eqref{Phiexpansion},  \eqref{Wexpansion} and  \eqref{Nexpansion}
into \eqref{lagrangian} we find 
\beq
\begin{split}
\label{action1}
\mathcal{L}_{\rm m}
 = &-\frac{1}{4}\mathrm{Re}f_{AB}\check{F}^A_{mn}\check{F}^{B\,mn}
+\frac{1}{8}\mathrm{Im}f_{AB}\epsilon^{mnpr}\check{F}^A_{mn}\check{F}^B_{pr}
-\frac{1}{16}\epsilon^{mnpr}B_{mn}e_A(\check{F}^A_{pr}+F^A_{pr})
\\
&+\frac12  \mathrm{Re}f_{AB} D^A D^B 
-\frac12 CD^A(e_A +2\mathrm{Im}f_{AB}m^B)
- \frac12 \mathrm{Re}f_{AB} m^Am^B C^2\\
&-\frac{i}{2}\,f_{AB}\lambda^A\sigma^m\partial_m\bar{\lambda}^B
-\frac{i}{2}\,f^*_{AB}\bar{\lambda}^A\bar{\sigma}^m\partial_m{\lambda}^B\\
&-\frac12 (ie_A +2 f_{AB}m^B) \eta\lambda^A-\frac12(-ie_A +2 f^*_{AB}m^B)\bar{\eta}\bar{\lambda}^A + \dots\ ,
\end{split}
\end{equation}
where we abbreviated the gauge invariant combination as
\beq
\check{F}_{mn}^A=F^A_{mn}-m^AB_{mn}\ ,
\eeq
and where the ellipses denote terms proportional to 
$\partial_i f_{AB}\,$.\footnote{%
The complete action can be found in \cite{schulgin}.}

The next step is to eliminate the auxiliary fields 
$D^A$
by their  equations of motion. Inserting the solution back into
\eqref{action1}  we arrive at 
\bea\label{result1}
\mathcal{L}_{\rm m}
& = &-\frac{1}{4}\mathrm{Re}f_{AB}\check{F}^A_{mn}\check{F}^{B\,mn}
+\frac{1}{8}\mathrm{Im}f_{AB}\epsilon^{mnpq}\check{F}^A_{mn}\check{F}^B_{pq}
-\frac{1}{16}\epsilon^{mnpq}B_{mn}e_A(\check{F}^A_{pq}+F^A_{pq}) -V\nn
\\
&&-\frac{i}{2}\,f_{AB}\lambda^A\sigma^m\partial_m\bar{\lambda}^B
-\frac{i}{2}\,f^*_{AB}\bar{\lambda}^A\bar{\sigma}^m\partial_m{\lambda}^B\\
&&-\frac12 (ie_A + 2 f_{AB}m^B)\, \eta\lambda^A- \frac12 (-ie_A + 2 f^*_{AB}m^B)\,\bar{\eta}\bar{\lambda}^A\ +\ \ldots\ ,\nn
\eea
where the  potential $V$ is given by 
\beq\label{potential1}
V =\frac12\,  \mathrm{Re}f_{AB} D^A D^B 
+ \frac12\, \mathrm{Re}f_{AB} m^Am^B\, C^2\ , \quad
D^A = \frac12\mathrm{Re}f^{-1AB}(e_B +2\mathrm{Im}f_{BC}m^C)C\ .
\eeq
We see that only part of the potential is generated by  $D$-terms.
In addition there is a contribution to the mass-term for the scalar $C$ 
which does not arise from eliminating an auxiliary field.
For $m^A =0$ the mass term for $C$ does come from a $D$-term and
is part of the Green-Schwarz term 
discussed earlier. Explicitly inserting $D^A$ we arrive at
\bea\label{potential}
V &=&\frac18\,\Big((e_A +2\mathrm{Im}f_{AC}m^C)\,\mathrm{Re}f^{-1AB}\,
(e_B +2\mathrm{Im}f_{BD}m^D)  + 4 \mathrm{Re}f_{AB} m^Am^B  \Big)\ C^2\nn  \\
&=& 
\frac18\,\Big((e_A -2i f_{AC}m^C)\,\mathrm{Re}f^{-1AB}\,
(e_B +2i\bar f_{BD}m^D)\Big)\ C^2\ .
\eea
The potential \eqref{potential}
can be viewed as the $N=1$ version of the $N=2$ Taylor-Vafa potential
given in \cite{TV} and rederived in \cite{LM}. Note that it does not follow from a superpotential
but instead from a different chiral superspace integral \eqref{lagrangian}.
In \cite{TGL} this potential has been 
derived by a Kaluza-Klein reduction of certain type IIB orientifolds.
One purpose of this paper was to show how this potential arises
from the  superspace action \eqref{lagrangian}.

Let us also discuss 
the mass terms of the antisymmetric tensor as they appear 
in the Lagrangian \eqref{result1}. They arise from
the couplings including  $\check F_{mn}$ and displaying them separately
one has (the analogous $N=2$ expression can be found in \cite{LM})
\bea\label{Bmass}
\mathcal{L}_{\rm mB} &=& -\frac14 M^2\, B_{mn} B^{mn} 
+ \frac18 M^2_T\, \epsilon^{mnpq} B_{mn} B_{pq}\ , \nonumber\\
  M^2&=&\mathrm{Re}f_{AB} m^A m^B\ ,\\
  M_T^2&=& \mathrm{Im}f_{AB} m^A m^B+\frac12 e_Am^A\ . \nonumber
\eea
For $m^A=0$ both mass terms vanish and one is left
with a massless antisymmetric tensor with a Green-Schwarz coupling
of the form $\epsilon^{mnpq} F_{mn} B_{pq}$ as can be seen from
\eqref{result1}.

\section{Dualities of an antisymmetric tensor}
In this section we discuss the dual formulations 
of the antisymmetric tensor.\footnote{%
This has also been discussed recently in \cite{AFL,DF}.} 
We perform the duality transformation
in superfields and then expand the dual action in components. 
Alternatively one can perform the duality transformation
at the component level
starting from \eqref{result1}. Of course, 
both methods give the same dual action
in components and the  details of the second route can 
be found in ref.\ \cite{schulgin}.

We need to distinguish between 
the massless case $m^A=0$ and the massive case $m^A\neq 0$.
A massless antisymmetric tensor is dual to a scalar or at the level
of superfields a linear multiplet is dual to a chiral multiplet.
On the other hand a massive antisymmetric tensor is dual
to a massive vector or at the superfield level a 
spinor superfield is dual to a massive vector multiplet.
Let us discuss both cases in turn.

\subsection{Pure Green-Schwarz term -- massless linear multiplet}
For  $m^A=0$ the Lagrangian can be expressed entirely in terms of $L$ and reads
\bea\label{Le}
{\cal L} &=& -\int d^2\theta d^2\bar{\theta} K(L)
+\frac{1}{4}\int d^2 \theta\Big(
f_{AB} W^AW^B + 2 e_A \Phi W^A\Big)\ +\ {\rm h.c.}\ ,\nonumber\\
&=& -\int d^2\theta d^2\bar{\theta} \big(K(L) -  e_A L V^A\big)
+\frac{1}{4}\int d^2 \theta f_{AB} W^AW^B \ +\ {\rm h.c.}\ ,
\eea
where we used \eqref{GS}.
In order to perform the duality transformation one treats 
$L$ as a real (but not linear) superfield and adds the term \cite{Quevedo,BGG}
\beq\label{deltaL}
\delta {\cal L} = \int d^2\theta d^2\bar{\theta}\, L(S+\overline S)\ ,
\eeq
where $S$ is a chiral superfield $\overline D_{\dot\alpha} S =0$.
The equations of motion for $S$ and $\overline S$ impose the constraint
\eqref{constraint} and lead back to the action \eqref{Le}.\footnote{%
The equations of motion for $S$ are found by representing $S$ in terms
of an unconstrained superfield $X$ via 
$S=\overline D^2 X$ and varying the action with respect to $X$.
(See \cite{BGG} for more details.)}
The equation of motion for $L$ on the other hand reads
\beq\label{LSrel}
-\partial_L K + e_A V^A + S+ \overline S = 0\ ,
\eeq 
which expresses $L$ in terms of the combination
$e_A V^A + S+ \bar S$. The precise relation depends on
the specific form of $K(L)$.
Inserting the solution of \eqref{LSrel}
into the sum of \eqref{Le} and \eqref{deltaL} one obtains
\beq\label{Ldual}
{\cal L} = \int d^2\theta d^2\bar{\theta}\, \hat K(e_AV^A + S+ \overline S)
+\frac{1}{4}\int d^2 \theta f_{AB} W^AW^B \ +\ {\rm h.c.}\ ,
\eeq
where  $\hat K = - K + (e_A V^A + S+ \overline S)L$ 
is the Legendre transform of $K(L)$.
The $U(1)$ gauge invariance of the vector multiplets
is maintained as long as $S$ transforms according to
\beq
V^A\to V^A + \Sigma^A +\overline \Sigma^A \ , \qquad 
S\to S - e_A\Sigma^A\ ,
\eeq
where $\Sigma$ is a chiral superfield $ \overline D_{\dot\alpha} \Sigma =0$.
We see that
$S$ can be absorbed into $V^A$ rendering one 
linear combination of the $V^A$  massive 
with a mass given by the second derivative of $\hat K$.
Thus $S$ is the (charged) Goldstone multiplet `eaten' by 
a linear combination of the $V^A$.

In order to derive the component form of the action \eqref{Ldual}
we continue to expand $V^A$ in a Wess-Zumino gauge with component
fields $(v^A_m, \lambda_\alpha^A, D^A)$ as specified in
\eqref{Wexpansion} but include the components of $S$ as physical
fields with a $\theta$-expansion
$S=E+\sqrt{2}\theta\psi + \theta\theta F$.
Inserted into \eqref{Ldual} we arrive at
\bea\label{Lint}
\mathcal{L}&=&
\frac{1}{2} \hat K' e_A D^A \ 
+\frac12 \hat K^{\prime\prime}\Big(2FF^*-\partial_m(\RE E)\partial^m(\RE E) \nonumber \\
&-& \frac{1}{2} \left(e_A v^A_m \ +\ 2 \partial_m (\IM E)\right)\left(e_B v^{B\,m}\ +\ 2\partial^m (\IM E)\right) \nonumber \\  
&+& i\sqrt{2} e_A (\psi \lambda^A  - \bar{\psi } \bar{\lambda}^A) -
i(\psi \sigma^m\partial_m\bar{\psi }+\bar{\psi
}\bar{\sigma}\partial_m\psi) \Big) \ 
+\ \tfrac14\hat K^{\prime\prime\prime\prime}\psi \psi \bar{\psi }\bar{\psi } \\
&+& \frac{1}{2}\Kh^{\prime\prime\prime} \Big(-F^*\psi \psi -F\bar{\psi }\bar{\psi }+\bar{\psi }\bar{\sigma}^m\psi\, 
(e_Av^A_m+2\partial_m(\IM E)\Big) \nonumber\\
&-&\frac{1}{4}\mathrm{Re}f_{AB}F^{A\,mn}F^B_{mn}+\frac{1}{2}\mathrm{Re}f_{AB}D^AD^B+\frac{1}{8}\IM f_{AB}\epsilon^{mnpr}F^A_{mn}F^B_{pr}\nn\\
&-&\frac{i}{2}\left(f_{AB}\lambda^A\sigma^m\partial_m\bar{\lambda}^B-{f}^*_{AB}\partial_m\lambda^A\sigma^m\bar{\lambda}^B\right)
+ \ldots \ ,\nn
\eea
where the ellipses once more denote terms proportional to $\partial_i f_{AB}$
and we abbreviated $\hat K^\prime = \partial_{\RE E} \hat K, etc.\,$.
We see that the real scalar $\IM E$ plays the role of the Goldstone boson
which can be absorbed into the linear combination 
$e_A v^A_m$ by an appropriate gauge transformation (unitary gauge)
leaving a mass term behind.

The auxiliary fields $F$ and $D^A$ can be eliminated by their 
equations of motions.  Neglecting fermionic contributions we find
\beq\label{FD}
F= 0 \ , \qquad 
D^B=-\frac{1}{2}\Kh^\prime\, e_A (\RE f)^{-1\,AB} \ .
\eeq
Inserted back into \eqref{Lint} and going to the unitary gauge
the bosonic action reads
\bea\label{Lbosonic}
\mathcal{L}_{\rm b}&=& 
-\frac{1}{4}\mathrm{Re}f_{AB}F^{A\,mn}F^B_{mn}+\frac{1}{8}\IM f_{AB}\epsilon^{mnpr}F^A_{mn}F^B_{pr}-
\frac{1}{2} e_A e_B v^A_m  v^{B\,m}\nn\\
&& -\frac12 \hat K^{\prime\prime}\partial_m(\RE E)\partial^m(\RE E) - V\ ,
\eea
where the potential is given by
\beq
V\ =\ \frac{1}{2}\,\RE f_{AB}D^AD^B = \frac18\, (\Kh^\prime)^2 \,
e_A (\RE f)^{-1\,AB}e_B\ .
\eeq
We see that this potential agrees with the dual potential 
given in \eqref{potential} for $\Kh^\prime=C$ which is just expressing
the Legendre transformation between the dual variables discussed below 
\eqref{Ldual}. As a consequence also the kinetic 
terms of the scalars in \eqref{Lkincomponent} and 
\eqref{Lbosonic} agree.

\newpage
\subsection{Duality for a massive tensor multiplet}
Let us now discuss the situation when $m^A \neq 0$ 
and thus the antisymmetric is massive (c.f.\ \eqref{Bmass}).
In this case $B_{mn}$ is dual to a massive vector or, 
at the level of superfields,
$\Phi_\alpha$ is dual to a massive vector multiplet. 

We start from the (first order) Lagrangian \cite{LR}
\begin{equation}\label{first.order}
\begin{split}
\mathcal{L}\ =& \ 
\int d^2\theta d^2\bar{\theta}\,
\Big(U(V^{0}) -  V^{0}(D\Phi+\bar D\bar\Phi) \Big)\ + {\mathcal
  L}_m\ ,
\end{split}
\end{equation}
where ${\mathcal L}_m$ is given in \eqref{lagrangian}.
$\Phi_\alpha$ continues to be a chiral spinor superfield while $V^{0}$ is 
a new real superfield in addition to the $n_V$ vector fields $V^A$ already
present in ${\mathcal L}_m$. As we will see shortly
$V^{0}$ will play the role of the dual massive vector 
multiplet. Note that \eqref{first.order} is not gauge invariant and therefore
we cannot take $V^0$ in the WZ-gauge.

Let us first check that eliminating $V^0$ by its equations of motion 
leads us back to the action ${\mathcal L}_{\rm kin}+{\mathcal L}_{\rm m}$
given in \eqref{Lkin} and \eqref{lagrangian} respectively.
{}From \eqref{first.order} we determine the field equation of $V^0$
to be  
\begin{equation}
\frac{\partial U}{\partial V^0}=2L=D\Phi+\bar D\bar\Phi\ .
\end{equation}
For appropriate functions $U(V^0)$ 
this can be solved for $V^0$ in terms of  $L$ 
\begin{equation}\label{umkehrung}
V^0=h(L)\ ,
\end{equation}
where $h(L)$ is the inverse function of $U'$.
Inserted back into (\ref{first.order}) we indeed obtain 
\begin{equation}
{\mathcal L}=-\int d^2\theta d^2\bar{\theta} K(L)\ +\ 
{\mathcal L}_{\rm m}\ ,
\end{equation}
where $K(L)=2h(L) L-U(h(L))$ is the Legendre transform of $U$.
This establishes that \eqref{first.order} is the correct
starting point for deriving the dual action to which we turn now.

The dual action can be obtained by eliminating $\Phi_\alpha$ in favor of $V^0$
using the equation of motion of $\Phi_\alpha$. 
More precisely we first replace $\Phi_\alpha$ by an 
unconstrained superfield $Y$ via 
$\Phi_{\alpha}=\overline D^2 D_{\alpha} Y$
and then vary the action \eqref{first.order} with respect
to $Y$. This yields
\beq\label{eom}
D^\alpha \Big((m^Af_{AB}m^B+\frac{i}{2}e_Am^A)\Phi_\alpha\Big)
=-\frac{1}2 D^\alpha\Big(W^0_\alpha + (im^Af_{AB}-\frac{1}{2}e_B)W^B_\alpha \Big)\ ,
\eeq
where $W^0= -\frac14 \overline D^2 D_\alpha V^0$.
The solution of \eqref{eom} reads
\beq \label{phi1}
\Phi_\alpha\ =\ -\frac{1}{2}\, \frac{(im^Af_{AB}-\frac{1}{2}e_B)W^B_\alpha
+W^0_\alpha}{m^C f_{CD}m^D+\frac{i}{2}e_Cm^C}\ \ .
\eeq
As promised $\Phi_\alpha$ is expressed in terms of 
$W^A_\alpha$ and $W^0_\alpha$ which is
the field strength of the newly introduced vector multiplet $V^0$.
Inserted  back into (\ref{first.order}) we obtain the Lagrangian 
for $n_V+1$ vector multiplets
\beq\label{Lvector}
\mathcal{L}=
\frac{1}{4}\int d^2\theta \hat{f}_{\hat{A}\hat{B}}W^{\hat{A}}W^{\hat{B}} 
+ \int d^2\theta d^2\bar{\theta}\, U(V^{0}) \ , 
\qquad \hat{A}=0, \ldots, n_V\ ,
\eeq
where we have introduced the $(n_V+1)\times(n_V+1)$ 
dimensional gauge coupling matrix $\hat{f}_{\hat{A}\hat{B}}$ given by
\beq\begin{split}\label{fhatdef}
\hat{f}_{AB}&=f_{AB}+\hat{f}_{00}^{-1}\,\hat{f}_{0A}\hat{f}_{0B}\ ,
\qquad \hat{f}_{0A}=\hat{f}_{00}\,(im^Bf_{AB}-\frac12 e_A)\ ,\\
\hat{f}_{00}&=[M^2+iM^2_T]^{-1}=[m^Af_{AB}m^B +\frac{i}2e_Am^A]^{-1}
\ .
\end{split}\eeq
($M$ and $M_T$ are defined in \eqref{Bmass}.)
We immediately see that due to the second term in \eqref{Lvector}
$V^0$ is a massive vector multiplet.
However, we have not yet fixed the original two-form
gauge invariance displayed in \eqref{Vtrans} and as a consequence
the action \eqref{Lvector} contains one unphysical vector multiplet.
This gauge invariance 
can be used to gauge away one of the original $n_V$ vector multiplets
leaving one massive and $n_V-1$ massless vector multiplets as the 
physical spectrum. (Note that $V^0$ is invariant under the
transformation \eqref{Vtrans} and thus cannot be gauged away.)

The fact that one of the vector multiplets in \eqref{Lvector} is unphysical
can also be seen from the gauge coupling matrix
$\RE \hat{f}_{\hat{A}\hat{B}}$ which has a zero eigenvalue while the matrix
of $\theta$-angles given by $\IM \hat f _{\Ah\Bh}$ has one constant eigenvalue.
This follows directly from \eqref{fhatdef} which using \eqref{Bmass} implies
\beq
m^A\hat{f}_{AB}=-\frac{i}{2}e_B\ ,\qquad m^A\hat{f}_{0A}=i\ .
\eeq 
Or in other words
$m^A\RE \hat{f}_{AB}=0$ and hence 
$\RE \hat f _ {\Ah\Bh}$ has the null vector $(0,m^A)$. 
This implies that one (linear combination) of the vector fields 
only has a topological coupling but no kinetic term.\footnote{%
This is again in complete analogy to the situation found in $N=2$ \cite{LM}.}

Let us give the Lagrangian \eqref{Lvector}
in components.
In order to do so 
we expand $V^0$ as
\bea\label{V0}
V^0 &=& \frac12 A^0+\sqrt 2 \theta \psi^0+\sqrt 2\bar\theta\bar\psi^0 -\theta\sigma^m\bar\theta v_m^0+\theta\theta F^0+\bar\theta\bar\theta F^{*0}\\
&&+i\theta\theta\bar\theta(\bar\lambda^0+\frac{1}{\sqrt 2}\bar\sigma^m\partial_m\psi^0)-i\bar\theta\bar\theta\theta(\lambda^0-\frac{1}{\sqrt 2}\sigma^m\partial_m\bar\psi^0)+\frac12\theta\theta\bar\theta\bar\theta(D^0+\Box A^0)\ ,\nn
\eea
where $A^0$ is real. 
Together with \eqref{Wexpansion} we insert \eqref{V0} into \eqref{Lvector} 
and arrive at
\beq\label{Lcomponent}
\begin{split}
\mathcal{L} = & -\frac14 \RE f_{\Ah\Bh} F^{\Ah\, mn} F^{\Bh} _{mn}
+ \frac{1}{8} \IM f_{\Ah \Bh} \epsilon^{mnpq} F^{\Ah}_{mn} F^{\Bh}_{pq}
- \frac14 U^{\prime\prime} v_m^0 v^{0m} \\
& 
+U^{\prime\prime}\big(F^0F^{0*}
-\frac{1}{2}\partial_m A^0 \partial^m A^0 \big)
-\frac{1}{2} U^\prime D^0-\frac12 \RE \hat{f}_{\hat{A}\hat{B}}D^{\hat{A}}D^{\hat{B}}\\
&-\frac{i}{2}\Big(\hat f _{\Ah\Bh}\lambda^\Ah\sigma^m\partial_m\bar\lambda^\Bh+\hat f^*_{\Ah\Bh}\bar\lambda^\Ah\bar\sigma^m\partial_m\lambda^\Bh\Big) \\
&+\frac12 U^{\prime\prime}\big(i\sqrt{2}( \psi^0 \lambda^0-\bar{ \psi^0 }\bar{\lambda}^0)-i( \psi^0 \sigma^m\partial_m\bar{ \psi^0 }+\bar{ \psi^0 }\bar{\sigma}\partial_m \psi^0) \big)
+\frac{1}{4}U^{\prime\prime\prime\prime} 
\psi^0  \psi^0 \bar{ \psi^0 }\bar{ \psi^0 }\\
&+\frac{1}{2}U^{\prime\prime\prime} \big(-F^{0*} \psi^0  \psi^0 
-F^0\bar{ \psi^0 }\bar{ \psi^0 }+\bar{ \psi^0 }\bar{\sigma}^m \psi^0 v_m^0\big)+\ldots\ ,
\end{split}
\eeq
where again we are neglecting terms proportional to 
$\partial_i \hat f_{\hat A\hat B}$.

The next step is to eliminate the auxiliary fields $D^{\hat A}$ and $F^0$.
$F^0$ can by straightforwardly eliminated by its equation of motion 
\beq\label{Feq}
F^0= \frac12 \frac{U^{\prime\prime\prime}}{U^{\prime\prime}}\, \bar\psi^0\bar\psi^0\ .
\eeq
For $D^{\hat A}$ the situation is more involved 
since one of the vector multiplets is unphysical and its 
equation of motion cannot be used.\footnote{%
This `difficulty' can also be
seen from the fact that 
$\RE \hat{f}_{\hat{A}\hat{B}}$ has a zero eigenvalue. }
Instead we should only vary with respect to the auxiliary fields
of the physical multiplets or in other words first fix the gauge invariance
\eqref{Vtrans}.

One way to obtain the appropriate constraint is to consider the
transformation properties of eq.\ \eqref{phi1}
under the transformation \eqref{Eichung}, \eqref{Vtrans}.
In section~\ref{superm}
we showed that the gauge transformation \eqref{Eichung}
can be used to go to a WZ-gauge for $\Phi$. In order to fix 
the corresponding gauge on the 
the right  hand side of \eqref{phi1} 
we first rewrite \eqref{phi1} as
\beq\label{phi3}
\begin{split}
\Phi_\alpha=&-\frac12 \frac{M^2}{M^4+M^4_T}\bigg(-(m^A\IM f_{AB}+\frac12
e_B)W^B_\alpha+W^0_\alpha+\frac{M^2_T}{M^2}m^A\RE f_{AB}W^B_\alpha\bigg)\\
&-\frac12\frac{iM^2_T}{M^4+M^4_T}\bigg((m^A\IM f_{AB}+\frac12 e_B)W^B_\alpha-W^0_\alpha+\frac{M^2}{M^2_T}m^A\RE f_{AB}W^B_\alpha\bigg)\ .
\end{split}
\eeq
The constraint on the $D$-terms arises from the $\theta$-component
of this equation. In the WZ-gauge one has (c.f.\ \eqref{Phiexpansion})
\beq
W^A|_{\theta}={\delta_{\alpha}}^{\beta}D^A
-\frac{i}{2}{(\sigma^m\bar{\sigma^n})_{\alpha}}^{\beta}F_{mn}^A\ ,\qquad 
\Phi|_{\theta}=-{\delta_{\alpha}}^\beta\frac12 C-\frac14
     {(\sigma^m\bar{\sigma}^n)_{\alpha}}^{\beta}B_{mn}\ .
\eeq
Inserted into \eqref{phi3} we see that the imaginary part 
of the right hand has to  vanish in the WZ-gauge or in other words
the constraint
\beq\label{Dconstraint}
(m^A\IM f_{AB}+\frac12 e_B)D^B-D^0+\frac{M^2}{M^2_T}\RE f_{AB}m^A D^B=0\ 
\eeq
has to be imposed.

Using the  constraint \eqref{Dconstraint} 
we can eliminate for example the $D^0$ field from
the Lagrangian \eqref{Lcomponent}
and are left with the following terms containing $D^A$
\beq\label{Dint}
\begin{split}
{\cal L}=&-\frac{1}{2} U^\prime 
\Big(\frac{M^2}{M_T^2}\RE f_{AB}m^A+m^A\IM f_{AB}+\frac{1}{2}e_B\Big)D^B\\
&-\frac{1}{2}\Big(\RE f_{AB}+\frac{M^2}{M^4_T}m^Cm^D\RE f_{AC}\RE
f_{BD}\Big)D^AD^B + \ldots\ .
\end{split}
\eeq
{}From \eqref{Dint} we can now determine 
the equation of motion for $D^A$ as 
\beq\label{Deq}
D^A=-\frac{1}{4}\, U^\prime\ \RE f^{-1\, AC}(e_C+2\IM f_{BC}m^B)\ .
\eeq
Inserting everything  
back into \eqref{Dint} then results in 
\beq\label{Lcomponentp}
\begin{split}
\mathcal{L} = & -\frac14 \RE f_{\Ah\Bh} F^{\Ah\, mn} F^{\Bh} _{mn}
+ \frac{1}{8} \IM f_{\Ah \Bh} \epsilon^{mnpq} F^{\Ah}_{mn} F^{\Bh}_{pq}
- \frac14 U^{\prime\prime} v_m^0 v^{0m} \\
& -\frac{1}{2}U^{\prime\prime}\partial_m A^0 \partial^m A^0 
-\frac{i}{2}\Big(\hat f _{\Ah\Bh}\lambda^\Ah\sigma^m\partial_m\bar\lambda^\Bh+\hat f^*_{\Ah\Bh}\bar\lambda^\Ah\bar\sigma^m\partial_m\lambda^\Bh\Big) \\
&+\frac12 U^{\prime\prime}\big(i\sqrt{2}( \psi^0 \lambda^0-\bar{ \psi^0 }\bar{\lambda}^0)-i( \psi^0 \sigma^m\partial_m\bar{ \psi^0 }+\bar{ \psi^0 }\bar{\sigma}\partial_m \psi^0) \big)\\
&+\frac{1}{4}\big(U^{\prime\prime\prime\prime} 
- \frac{U^{\prime\prime\prime}}{U^{\prime\prime}}\big)\,
\psi^0  \psi^0 \bar{ \psi^0 }\bar{ \psi^0 }
+\frac{1}{2}U^{\prime\prime\prime} \bar{ \psi^0 }\bar{\sigma}^m \psi^0 v_m^0 - V \ +\ \ldots\ ,
\end{split}
\eeq
where the potential is given by 
\bea
V&=&\frac{1}{2}\RE f_{AB} D^A D^B +\frac18 M^2 U^{\prime 2} \\
&=&\frac{1}{32}U^{\prime 2}\bigg(\left(e_A+2\IM f_{AC}m^C\right)\RE f^{-1\, AB}\left(e_B+2\IM f_{BD}m^D\right)+4\RE f_{AB}m^Am^B\bigg) \nn\ .
\eea
This potential
indeed coincides with the potential \eqref{potential}
for $\frac12 U^{\prime}(A) = C$. For this identification 
also the kinetic terms agree which follows from the fact that $U(A)$
and $K(L)$ are related by a Legendre transformation.
Note that \eqref{Lcomponentp} contains apart from the kinetic terms
a mass term for the vector
$v_m^0$ and its fermionic partners $\psi^0, \lambda^0$ which is proportional
to $U^{\prime\prime}$.
This ends our discussion of the dual action and 
we showed that in both formulations the scalar potential agrees.

Let us briefly summarize. Motivated and guided by the results of 
refs.\ \cite{LM,TGL} we proposed an $N=1$ superfield action \eqref{Lkin},
\eqref{lagrangian} for a massive tensor multiplet coupled to
$n_V$ vector and $n_c$ chiral multiplets. We computed the component
form of the action and showed that the resulting potential \eqref{potential1},
\eqref{potential}
agrees with the potential obtained in a Kaluza-Klein reduction
of type IIB orientifolds. The potential has the `unusual' feature
that it is not solely determined  by auxiliary fields but has
a direct mass term for the scalar in the tensor multiplet.
We also discussed the dual superspace action where the massive tensor multiplet
has been dualized to a massive vector multiplet.
We computed the component form of the dual action 
and showed that the potential in the two dual formulation coincide.

\vskip 1cm

\subsection*{Acknowledgments}

This work is supported by DFG -- The German Science Foundation,
the European RTN Programs HPRN-CT-2000-00148 and the
DAAD -- the German Academic Exchange Service.

We have greatly benefited from conversations with
Thomas Grimm and Florian Schwennsen.

\newpage


\end{document}